\documentclass[preprint]{elsarticle}
\usepackage{hyperref,graphicx,amsmath,amssymb,xcolor,leftidx,bm}
\journal{International Journal of Non-Linear Mechanics}
\newcommand{\Sp}{S_\parallel}
\newcommand{\So}{S_\perp}
\newcommand{\tb}{\tilde\beta}
\newcommand{\tg}{\tilde\gamma}
\newcommand{\tk}{\tilde\kappa}
\newcommand{\trh}{\tilde\rho}

\newcommand{\be}{\bm{e}}
\newcommand{\bex}{\bm{e}_x}
\newcommand{\bey}{\bm{e}_y}
\newcommand{\bez}{\bm{e}_z}
\newcommand{\ber}{\bm{e}_r}
\newcommand{\bet}{\bm{e}_\Theta}
\newcommand{\bffa}{\bm{f}}
\newcommand{\bfpa}{\bm{f}_\perp}
\newcommand{\bffb}{\bm{h}}
\newcommand{\bfpb}{\bm{h}_\perp}

\newcommand{\bn}{\bm{n}}
\newcommand{\bt}{\bm{t}}
\newcommand{\by}{\bm{y}}
\newcommand{\bA}{\bm{A}}
\newcommand{\bAe}{\bm{A}_\text{e}}
\newcommand{\bAg}{\bm{A}_\text{g}}

\newcommand{\bC}{\bm{C}}
\newcommand{\bCe}{\bm{C}_\text{e}}

\newcommand{\bEe}{\bm{E}_\text{e}}
\newcommand{\bF}{\bm{F}}
\newcommand{\bFg}{\bm{F}_\text{g}}
\newcommand{\bFe}{\bm{F}_\text{e}}

\newcommand{\bI}{\bm{I}}
\newcommand{\bIs}{\bm{I}_\text{s}}
\newcommand{\bK}{\bm{K}}

\newcommand{\bN}{\bm{N}}
\newcommand{\bT}{\bm{T}}
\newcommand{\bY}{\bm{Y}}
\newcommand{\bna}{\bm{\nabla}}
\newcommand{\bzero}{\bm{0}}

\newcommand{\mbC}{\mathring{\bm{C}}}

\newcommand{\mbF}{\mathring{\bm{F}}}

\newcommand{\cB}{\mathcal{B}}
\newcommand{\cE}{\mathcal{E}}
\newcommand{\cF}{\mathcal{F}}

\newcommand{\fJ}{\mathfrak{J}}
\newcommand{\fJg}{\mathfrak{J}_\text{g}}
\newcommand{\Jg}{J_\text{g}}

\newcommand{\bchi}{\boldsymbol{\chi}}

\newcommand{\tr}{\mathop{\mathrm{tr}}}
\newcommand{\curl}{\mathop{\mathrm{curl}}}

\newcommand{\hypgeo}[2]{\leftidx{_#1}{F}{_#2}}

\begin{document}

\title{Encasement as a morphogenetic mechanism:\\ The case of bending\tnoteref{t1}}
\tnotetext[t1]{Dedicated to Martine Ben Amar whose profound researches have been crucial in illuminating the interplay between Mechanics and Biology.}

\author[pm]{Paolo Biscari\corref{cor1}}
\ead{paolo.biscari@polimi.it}
\author[pm]{Pier Luigi Susini}
\ead{pier.susini@mail.polimi.it}
\author[pd]{Giovanni Zanzotto}
\ead{giovanni.zanzotto@unipd.it}
\cortext[cor1]{Corresponding author}

\address[pm]{Department of Physics, Politecnico di Milano. Piazza Leonardo da Vinci 32, 20133 Milano, Italy}
\address[pd]{DPG, Universit\`a di Padova. Via Venezia 8, 35131 Padova, Italy}

\begin{abstract}
We study how the encasement of a growing elastic bulk within a possibly differently growing elastic coat may induce mechanical instabilities in the equilibrium shape of the combined body. The inhomogeneities induced in an incompressible bulk during growth are also discussed. These effects are illustrated through a simple example in which a growing elastic cylinder may undergo a shape transition towards a bent configuration.
\end{abstract}


\maketitle

\section{Introduction}\label{sec:intro}

Shape is one of the most prominent features of cells, plants, organs, and of living beings in general. Specific shapes may play a crucial role in enabling the achievement of specialized biological goals, which might allow for an organism's survival and thriving in its environment. Morphogenesis has been linked to mechanical origins in a natural way because mechanics offers innumerable examples of systems whose equilibrium configurations depend on external parameters, with bifurcations among equilibria which may imply very different shapes and structures for a body. Perhaps the most studied and exploited example of such morphogenetic bifurcation phenomena is given by Euler's elastica. In more recent times we mention the studies by Biot \cite{63biot} who understood how three-dimensional rubbers under compression may develop surface instabilities, as have also later been found in soft, strain-hardening materials \cite{10bisoma, 06gordesben}.

A mechanical approach in the creation of form for bodies undergoing growth was envisaged already at the beginning of last century by Thompson \cite{GrowthForm}. Interesting studies along these lines were later performed, such as for instance \cite{81conbr}, where shell theory was used to estimate the surface stresses that can be induced in a growing surface, showing how they may related to splitting and cracking in vegetables. These methods give very fertile ground for the mechanical investigation of bio-morphogenesis, greatly expanding the scope of continuum mechanics well into the boundaries of biophysics and biology in general. A growing body indeed evolves in a natural way under the effect of a number of external parameters which vary with time inducing morphogenetic effects with lesser need for genetic encoding of information. This offers attractive avenues for the explanation of form  \cite{oulesgenes}, within a background of natural hypotheses of genetic parsimony.

Understanding in terms of mechanical instabilities has since been proposed for numerous observed shapes of growing bodies, using a wide variety of approaches, and sometimes exhibiting remarkable agreement with experimental observations. To name but a few examples, buckling under external stress has been related to morphogenesis for instance in growing spheroidal shells \cite{08YinPNAS}, or in the growth of other constrained systems \cite{11MartineJMPS, 05goben}. More complex shapes, observed in plant organs such as long leaves \cite{09lima}, or in blooming flowers \cite{benamarlibro, 11lima}, have also been modelled by adapting the theory of elastic shells to a growing surface. See \cite{12LiCaoSM} for a review of the mechanics of buckling-related morphogenesis. Further recent studies have further analyzed the growth of vegetable matter, as in growing pumpkins \cite{11hurial} or in the ripening of kiwi fruit \cite{13HallJEB}, see also the review \cite{11mida}. In the biomedical sciences, the investigation of mechanical instabilities has helped investigate the role of external stress in tumor growth \cite{02ambmol,02MMMASambrosipreziosi}, and mechanics has since provided fruitful soil for cancer modelling \cite{09pretos}. Recent work has also related mechanical instabilities to morphogenetic development during the growth of animal organs \cite{mahagut11, 14taber}, and in brain formation \cite{14budJMPS, 11hohmah, brainmaha14}.

In most cases, the basic trigger for morphogenetic instabilities derives from the presence of differential growth and distinct mechanical properties in different neighboring parts of a growing body. Thus, adjacent layers or domains sharing open two-dimensional \cite{12lucantonio,Jin14} or one-dimensional \cite{mahagut11} adhesion boundaries have been examined. Confinement of a growing body within a closed regular two-dimensional boundary not undergoing growth has also been considered \cite{Li11, wang15}. Less explored appear to be the morphogenetic possibilities for {\it encased} growing bodies, wherein a growing bulk is surrounded/protected by a closed regular growing surface layer (as with the skin or rind of fruits), whose mechanical and growth properties will in general differ from those in the bulk.

The aim of the present study is thus to highlight how encasement may generate interesting morphogenetic instabilities in growing bodies. We treat growth by following the approach introduced in \cite{02dicqui,94rod}, and extensively adopted in theoretical studies of growing systems, as in \cite{05bengor,08benamarPRL} among many examples. A peculiarity in our approach is that two multiplicative decompositions for the deformation gradient must be introduced to correctly account for the possibly independent growth of the bulk and the coat. In the bulk, growth is described by the factor $\bFg$ in the decomposition $\bF=\bFe\bFg$ of the three-dimensional deformation gradient $\bF$. Simultaneously, a similar decomposition $\bA=\bAe\bAg$ is introduced for the surface deformation gradient $\bA$. The strain energy then depends only on the elastic factors $\bFe$ and $\bAe$. By following \cite{75gumo,99stog} in our treatment of elastic continua surrounded by elastic surfaces, we consider an incompressible neo-Hookean material for the bulk, and F\"oppl-Van K\'arm\'an elastic coat, undergoing finite strains.

The growth components $\bFg,\bAg$ are at least partially determined by external processes, typically of biochemical origin. Furthermore, material remodeling may possibly influence such processes, and at least partially relieve the residual stresses which may have been originated by incompatible growth. The choice of including or not such an evolution for the growth components determines whether we are considering morphogenesis at stress-free conditions vs.\ in the presence of residual stresses. This might for instance depend on the relative time scales of growth and relaxation (see \cite{14bisdctur} for an example of how a soft material may exhibit purely elastic or apparently plastic behavior, depending on the deformation and relaxation times). We keep our study as simple as possible here, and do not consider any remodeling, so that in general the deformed shapes we characterize possess residual stresses (as, for instance, the systems considered by \cite{09lima}).

Our analysis is based on the combination of two effects. The first is the role that non-homogeneous growth may play in giving incompressible bodies the possibility of undergoing apparently non-isochoric deformations. The second factor is showing how encasement, i.e.~the coexistence of a growing bulk inside a possibly differently growing skin, may affect the equilibrium configurations of the body even in the absence of instabilities of the bulk or the boundary separately.

Here we illustrate these effects by means of a simplest example, that is, a bending instability for an isotropically growing straight cylinder encased within a cylindrical boundary which is in turn growing at possibly different rates in the longitudinal and transverse directions. Such growth anisotropy may occur, for instance, when parallel fibers are present in the skin, which may also possess anisotropic elastic moduli. We show how encasement induces, in a range of growth and elastic parameters, instabilities which lead to new equilibria with a bent shape for the cylinder. Depending on the material properties, other instabilities, which we do not explore here, may occur in encased growing cylinders, such as bulging, twisting, or others (see for instance the deformations considered in \cite{joanny12}). Our analysis shows that for a growing body the constraint of encasement is different from simple confinement as in \cite{Li11, wang15}, as the growth of the external coat along with the growth of the bulk creates instabilities which might not be present when the coat is (deformable but) incapable of growth.

This paper is organized as follows. In Section~\ref{sec:incgro} we show how inhomogeneous growth may provide new degrees of freedom for the distortion of incompressible growing bodies. In Section~\ref{sec:1} we analyze the effects of encasement on growth, showing how mechanical instabilities may arise when a growing body is surrounded by a differently growing coat. In Section~\ref{sec:bendin} we evidence a bending instability generated by encasement. An Appendix contains some of the computations leading to the results presented in the main text.

\section{Growth in incompressible materials}\label{sec:incgro}

In this section we analyze how inhomogeneous growth may enlarge the class of deformations available to incompressible materials. Let $\bchi$ be the map transforming the reference configuration into the present placement, and $\bF=\bna\chi$. We model growth by assuming \cite{02dicqui, 94rod} that $\bF$ may be decomposed as the product of an elastic and a growth component:
\begin{equation}\label{eq:solitadec}
\bF=\bFe\bFg.
\end{equation}
The growth tensor $\bFg$ describes how growth would locally shape body volume elements, were they allowed to grow stress-free, that is, in the absence of the surrounding elements. The balance equations driving $\bFg$ may not entirely be included in the bio-mechanical model. When this is the case, some or all of its entries may be explicitly specified, and therefore treated as external parameters. We remark that $\bFg$ need not be the gradient of any `growth deformation', as growth is a local phenomenon which may induce a lack of compatibility whenever $\curl\bFg\neq\bzero$. Moreover, since mass is not conserved during growth, no isochoricity condition must be \emph{a priori} enforced on $\det\bFg$.

The elastic distortion $\bFe$ is defined by the decomposition \eqref{eq:solitadec}: $\bFe=\bF\bFg^{-1}$. The assumption that $\bFg$ identifies the current stress-free configuration implies that the strain-energy density must be a function of the elastic strain
\begin{equation}\label{eq:defce}
\bCe=\bFe^\top\bFe=\bFg^{-\top}\bF^\top\bF\bFg^{-1}=\bFg^{-\top}\bC\bFg^{-1},
\end{equation}
where $\bC=\bF^\top\bF$ is the standard Lagrangian strain tensor. As $\bCe$ is itself a function of $\bC$, any strain-energy density depending on $\bCe$ automatically complies with frame-invariance requirements.

The incompressibility constraint establishes an isochoricity condition on the elastic distortion $\bFe$, so that
\begin{equation}\label{eq:detfg}
\det\bF=\det\bFg.
\end{equation}
If the point-wise mass supply is assigned, the map $\det\bFg$ must be treated as an external parameter, and incompressibility limits as usual the class of deformations available to the growing body. If, on the contrary, the system deforms sufficiently slowly, apparently non-isochoric deformations become possible, in which growth accommodate local volume variations by concentrating mass production at sites undergoing greater expansion $\det\bFg$. When this is the case, the incompressibility constraint applies only globally, and the determinant of $\bF$ may neither be 1, nor even uniform. Much more general deformations thus become available to a slowly growing body, the map $\det\bF$ giving the explicit details on the local mass production.

\section{Encasement}\label{sec:1}

\begin{figure}[t!]\centering
\begin{tabular}{ccc}
\includegraphics[width=17mm,clip=]{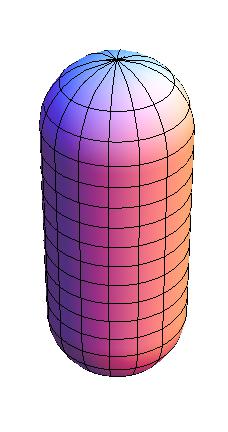} &\hspace{1mm} &
\includegraphics[width=42mm,clip=]{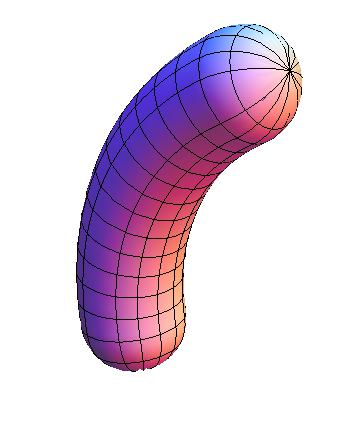} \\
\end{tabular}
\caption{(Left) Reference straight capped encased cylinder. (Right) Grown, stretched and bent configuration for the same cylinder.}
\label{fig:cylinders}
\end{figure}

We now study how the independent growth of an elastic bulk and of a closed, regular elastic surface surrounding it, may induce morphological instabilities. We consider the simplest geometry and constitutive assumptions on the material to better evidence the origin of such instabilities. The bulk of the body is given by a capped, isotropically growing cylinder made of simple incompressible neo-Hookean material, coated with a possibly anisotropic F\"oppl-von K\'arm\'an elastic surface undergoing finite strains. Two half spherical caps are considered, with centers on the cylinder axis and suitable radius, to enforce encasement, while ensuring $C^1$-regu\-larity of the external boundary, see Fig.~\ref{fig:cylinders}, left. We denote $r_0,h_0$ respectively the cylinder radius and height in the reference configuration, which in cylindrical coordinates is given by
\begin{equation}\label{eq:refconf}
\begin{aligned}
\cB_0&=\big\{P\in\cE:P=O+\rho\cos\Theta\,\bex+\rho\sin\theta\,\bey+Z\,\bez,\\
&\phantom{=\big\{}Z\in(-r_0,h_0+r_0),\,\rho\in\big[0,R_{r_0,h_0}(Z)\big),\,\Theta\in[0,2\pi)\big\},
\end{aligned}
\end{equation}
where $\{\bex,\,\bey,\,\bez\}$ is a fixed orthogonal basis and, for any positive $r$ and $h$, $R_{r,h}\colon (-r,h+r)\to \mathbb{R}^+$ is defined as
\begin{equation}\label{eq:r0z}
R_{r,h}(z)=
\begin{cases}
\sqrt{r^2-z^2} & \text{if } z\in(-r,0]\\
r & \text{if } z\in(0,h)\\
\sqrt{r^2-(z-h)^2} & \text{if } z\in[h,h+r).
\end{cases}
\end{equation}
The reference body is therefore identified by the reference radius $r_0$ and height $h_0$ of its cylindrical part.

As mentioned in the Introduction, we are interested in determining under which conditions the straight shapes of the capped cylinder become unstable, in particular in favor of bent stable shapes. We perform our analysis by introducing a parameterized family of shapes, and checking the stability of straight cylinder configurations against bending. The critical values we identify in this way thus provide a sufficient criterion for the loss of stability of the straight configuration. Other instabilities, not presently analyzed, could in principle occur at different critical thresholds, depending on the values of the constitutive parameters.

In detail, the family of bent shapes that we consider is obtained by assuming that the axis of the cylinder becomes the arc of a circle with curvature $\kappa$, as in the right panel of Fig.~\ref{fig:cylinders} (see the Appendix for further details):
\begin{align}
\cB_{\kappa}&=\big\{P\in\cE:P=Q_\kappa+\rho\cos\theta\,\bex+\big(\rho\sin\theta-\kappa^{-1}\big)(\cos\kappa z\,\bey-\sin\kappa z\,\bez),\notag\\
&\phantom{=\big\{}z\in(-r,h+r),\,\rho\in[0,R_{r,h}(z)),\,\theta\in[0,2\pi)\big\},\label{eq:defcyl}
\end{align}
where $Q_\kappa=O+\kappa^{-1}\,\bey$ is the center of the circle on which the axis lies. The domain $\cB_{\kappa}$ approaches the upright shape $\cB_0$ for vanishing $\kappa$.

The encased cylinder undergoes growth, with details on the bulk and surface growth distortions $\bFg,\bAg$ provided below, while it deforms to achieve the current configuration $\cB_\kappa$, determined by minimizing the strain energy below.

\begin{itemize}
\item The bulk material undergoes a possibly inhomogeneous dilation, characterized by the growth tensor (see \eqref{eq:detfg})
\begin{equation}
\bFg\big|_\text{bulk}=(\det\bF)^{1/3}\bI,
\end{equation}
where $\bI$ is the identity. Since the bulk material is assumed to be incompressible, growth fixes the total volume of the current configuration . We therefore introduced a positive scalar $\alpha$ which represents the relative volume increase in the grown body, and require
\begin{equation}\label{eq:currvol}
V_\text{curr}=\alpha^3V_0.
\end{equation}

\item The coat of the cylinder undergoes an independent growth process, described by anisotropic elongation factors $\beta,\gamma$, respectively in the longitudinal and radial directions in the cylindrical part of the coat, plus an isotropic growth factor $\gamma$ for the semi-spherical caps. We do not require the coat material to be inextensible, although stretches with respect to the locally grown configuration will be energetically penalized.
\end{itemize}

In this framework, growth is therefore encoded in the geometrical dilation factors $\alpha,\beta,\gamma$, fixed by external processes, which we treat as parameters in the analyzed solutions of the problem. We remark that encasement induces a lack of compatibility in the field $\bFg$ whenever $\beta$ differs from $\gamma$, as no dilation of the growing bulk may succeed in accommodating anisotropic surface growth. This occurs often in growth processes, and is at the basis of the existence of residual stresses~\cite{94rod}.

\subsection{Strain energy}

To determine the optimal deformed configuration, we follow \cite{99stog} and assume that the total strain energy of the encased cylinder can be decomposed into a bulk and a surface contribution
\begin{equation}\label{eq:stren}
\cF[\bchi]=\int_{\cB_0}\sigma(\bC)\,\Jg\,dV+\int_{\partial\cB_0}\varsigma(\bA,\bK)\,\fJg\,dA,
\end{equation}
in the absence of any loads. The strain-energy density depends in fact only on the elastic distortion but, by using \eqref{eq:defce} for the bulk term and a similar decomposition for the surface deformation gradient (see \eqref{app:isostrain} in the Appendix), it may be written as density in the reference configuration, depending on the Lagrangian strain $\bC$, the surface deformation gradient $\bA$, and the curvature tensor $\bK$. The Jacobians $\Jg,\fJg$, respectively of the bulk and surface growth transformations are included in \eqref{eq:stren} because the strain-energy density is defined with the respect to grown volume/area elements.

We assume a simple, isotropic neo-Hookean behavior for the bulk strain-energy density:
\begin{equation}\label{eq:defsigma}
\begin{aligned}
\sigma(\bC)&=\textstyle\mu\big(\tr\bCe-3\big)=\mu\Big(\tr\big(\bFg^{-\top}\bC\bFg^{-1}\big)-3\Big)\\
&=\textstyle\mu\Big(\tr\big((\det\bF)^{-2/3}\bC\big)-3\Big)=\mu\big(\tr\mbC-3\big)
\end{aligned}
\end{equation}
with $\mu$ the shear modulus, and $\mbC=(\det\bC))^{-1/3}\bC$ the isochoric strain.

To include in our study materials with possibly anisotropic (e.g., fiber-structured) coats \cite{84spe}, we choose the following anisotropic expression for the surface strain-energy density
\begin{equation}\label{eq:defvarsigma}
\varsigma(\bA,\bK)=\textstyle \Sp(\bEe\bT^Z\cdot\bT^Z)^2
+\So(\bEe\bT^\Theta\cdot\bT^\Theta)^2+B\,\fJ_\text{e}\,H^2
\end{equation}
where $\bT^Z,\bT^\Theta$ are respectively the longitudinal and azimuthal tangent unit vectors on the reference surface (see \ref{sec:appsurf}), $\Sp$ and $\So$ represent the longitudinal/transverse 2D stiffness moduli, and $B$ the bending stiffness. The first and second terms in the surface strain-energy (\ref{eq:defvarsigma}) depend on the elastic component $\bAe$ of the surface deformation gradient $\bA=\bAe\bAg$, as $\bEe=\bAe^\top\bAe-\bIs$, where $\bIs$ denotes the identity operator on the tangent planes to $\cB_0$. The term penalizing the mean curvature $H$ carries a further factor $\fJ_\text{e}=\det\bAe$, as it penalizes curvature in the current configuration. Eq.~\eqref{eq:defvarsigma} defines for the coat a finite-elasticity counterpart of the strain energy of a classical F\"oppl-von K\'arm\'an plate \cite{villaggio}.

\section{Bending instability}\label{sec:bendin}

We now examine under which conditions the straight configuration may become unstable, possibly leading the cylinder to undergo a morphological transformation towards a bent shape (some of the computations are given in the Appendix). We remark that our model treats separately the bulk and surface strains, so that fairly straightforward changes of the computations presented below could also account for boundary slip and/or detachment during encased-growth phenomena. However, these are absent in the ensuing first analysis of encasement instabilities, as the family of deformations which we consider now is continuous at the coat-bulk boundary.

We decompose the capped cylinder into the cylindrical part, plus the two spherical caps, and consider the family of deformations $\bchi$ explicitly defined in the Appendix (see \eqref{app:defcoor}), parameterized by the height and radius $h,r$ of the current configuration, and the curvature $\kappa$ of the cylinder axis. The deformation gradient of the bulk transformation in the cylindrical part of the material (see \eqref{app:defcoor}) may be written in the orthogonal basis used in \eqref{eq:refconf} as
\begin{equation}\label{eq:bulkF}
\begin{aligned}
\bF(X,Y,Z)&=\frac{r}{r_0}(\bex\otimes\bex+\bffa(Z)\otimes\bey)+\frac{h}{h_0}\left(1-\frac{\kappa r Y}{r_0}\right)\bfpa(Z)\otimes\bez,
\end{aligned}
\end{equation}
with
\begin{align}
\bffa(Z)&=\cos\frac{\kappa h Z}{h_0}\,\bey-\sin\frac{\kappa h Z}{h_0}\,\bez,\ \quad
\bfpa(Z)=\sin\frac{\kappa h Z}{h_0}\,\bey+\cos\frac{\kappa h Z}{h_0}\,\bez.
\end{align}
The Jacobian
\begin{equation}\label{eq:detF}
J=\det\bF=\frac{hr^2(r_0-\kappa r Y)}{h_0r_0^3}
\end{equation}
provides the volumetric dilation factor in the bulk, and sets a maximum admissible value for the curvature $\kappa$ for the deformation to be regular. More precisely, since $|Y|\leq r_0$, the deformation is regular only when $\kappa r<1$, so that
\begin{equation}\label{eq:kmax}
0\leq \kappa<r^{-1}.
\end{equation}

The cylindrical contribution to the volumetric strain-energy density gives (see \eqref{eq:defsigma} and \eqref{app:trmbC})
\begin{equation}\label{eq:volen}
\begin{aligned}
\sigma\big|_\text{cyl}&=\frac{\mu hr^2(r_0-\kappa r Y)}{h_0r_0^3}
\left[2\Big(\frac{h_0 r}{h(r_0-\kappa  rY)} \Big)^{\frac23} + \Big(\frac{h(r_0-\kappa  r Y)}{h_0 r}\Big)^{\frac43}-3\right].
\end{aligned}
\end{equation}
The caps only undergo a dilation by a factor $r/r_0$, so that $\mbC\big|_\text{caps}=\bI$, and $\sigma\big|_\text{caps}=0$.

Identifying the elastic component of the surface deformation gradient $\bAe$ and the mean curvature $H$ requires lengthier computations, reported in the Appendix. In the deformed configuration, the cylindrical section of the coat stores a strain energy given by (see \eqref{app:isostrain} and \eqref{app:H}):
\begin{equation}
\begin{aligned}
\varsigma\big|_\text{cyl}&=\frac{\Sp}{4}\left(\frac{h^2(1-\kappa r\sin\Theta)^2}{\beta^2 h_0^2}-1\right)^2+\frac{\So}{4}\left(\frac{r^2}{\gamma^2 r_0^2}-1\right)^2\\
&+\frac{Bh}{4h_0 r_0 r(1-\kappa r\sin\Theta)}.
\end{aligned}
\end{equation}
The coat caps undergo an isotropic extension of factor $r/r_0$, remaining half-spheres of mean curvature $H=-r^{-1}$ (see \eqref{app:Hcaps}). Therefore
\begin{align}
\varsigma\big|_\text{caps}&=\frac{(\Sp+\So)}{4}\left(\frac{r^2}{\gamma^2 r_0^2}-1\right)^2+\frac{ B}{r_0^2}.
\end{align}

Let us now introduce the dimensionless parameters $u=r/(\alpha r_0)$, $v=h/(\alpha h_0)$, which reflect whether the current shape respects the dimensions obtained by a simple rescaling of the reference shape with a factor dictated by \eqref{eq:currvol}. Let further $\tk=\alpha r_0\kappa$ measure the bending state of the current configuration, and $\tb=\beta/\alpha$, $\tg=\gamma/\alpha$ be the rescaled surface growth factors. We also introduce the reference aspect ratio $\Lambda_0=r_0/h_0$. Notice that ($u$, $v$, $\tk$) parameterize the current configuration, while the parameters ($\tb$, $\tg$) characterize the normalized anisotropic growth. The total strain energy takes the form
\begin{align}
&\cF=\int_{\cB_0}\sigma(\bFe)\,\Jg\,dV+\int_{\partial\cB_0}\varsigma(\bA,\bK)\,\fJg\,dA\\
&=\mu \alpha^3 h_0 r_0^2 u^2v\int_0^{1}\tilde\rho\,d\tilde\rho\int_0^{2\pi}d\Theta
\bigg[\frac{2u^\frac23(1-\tk u\sin\Theta)^\frac13}{v^\frac23}+ \frac{v^\frac43(1-\tk u \sin\Theta)^\frac73}{u^\frac43}-3\bigg]\notag\\
&+\alpha^2r_0h_0\tb\tg\int_0^{2\pi}\!\!d\Theta\bigg[\frac{\Sp}{4}\left(\frac{v^2(1-\tk u\sin\Theta)^2}{\tb^2 }-1\right)^2+\frac{\So}{4}\left(\frac{u^2}{\tg^2}-1\right)^2\bigg]\notag\\
&+\frac{B vh_0}{4ur_0}\int_0^{2\pi}\frac{d \Theta}{1-\tk u\sin\Theta}
+\pi \alpha^2r_0^2\tg^2(\Sp+\So)\left(\frac{u^2}{\tg^2}-1\right)^2+4\pi B.\notag
\end{align}

To compute the above integrals we define, for any $|a|<1$,
\begin{align}
f_1(a,b)&=\frac{1}{2\pi}\int_0^{2\pi}(1-a\sin\Theta)^b d\Theta= \textstyle \hypgeo{2}{1}\left(\frac12(1-b)\,,\,-\frac12b\,;\,1\,;\,a^2\right)
\end{align}
with $\hypgeo{2}{1}$ the Gauss hypergeometric function such that
\begin{equation}
\begin{aligned}
&f_1(a,0)=f_1(a,1)=1,\quad \text{and}\\
&f_1(a,b)=1+\textstyle\frac14a^2b(b-1)+O(a^4)\quad  \text{as }a\to 0.
\end{aligned}
\end{equation}
For future use, we also introduce, for any $|a|<1$,
\begin{align}
f_2(a,b)&=2\int_0^1\trh \,f_1(a \trh,b) \,d\trh= \textstyle \hypgeo{2}{1}\left(\frac12(1-b)\,,\,-\frac12b\,;\,2\,;\,a^2\right).
\end{align}
We thus obtain
\begin{align}
\cF&=2\pi\mu h_0 r_0^2\alpha^3 u^2v\int_0^{1}\tilde\rho
\bigg[2\frac{u^\frac23}{v^\frac23}f_1(\tk u\trh,\tfrac13)+
\frac{v^\frac43}{u^\frac43}f_1(\tk u\trh,\tfrac73)-3\bigg]d\tilde\rho\notag\\
&+2\pi\alpha^2r_0h_0\tb\tg\bigg[\frac{\Sp}{4}\left(\frac{v^4f_1(\tk u,4)}{\tb^4 }-\frac{2v^2f_1(\tk u,2)}{\tb^2 }+1\right)+\frac{\So}{4}\left(\frac{u^2}{\tg^2}-1\right)^2\bigg]\notag\\
&+\frac{\pi B vh_0f_1(\tk u,-1)}{2ur_0}
+\pi \alpha^2r_0^2\tg^2(\Sp+\So)\left(\frac{u^2}{\tg^2}-1\right)^2+4\pi B.\label{eq:oneint}
\end{align}
The hypergeometric integrals in \eqref{eq:oneint} may again be computed, and provide the strain energy per unit current volume ($\Psi=\cF/(\pi\alpha^3r_0^2h_0)$)
\begin{align}
\Psi&=\mu \big(2u^\frac83v^\frac13 f_2(\tk u,\tfrac13)+
u^\frac23v^\frac73f_2(\tk u,\tfrac73)-3u^2v\big)\notag\\
&+\frac{2\tb\tg}{\alpha r_0}\bigg[\frac{\Sp}{4}\left(\frac{v^4f_1(\tk u,4)}{\tb^4 }-\frac{2v^2f_1(\tk u,2)}{\tb^2 }+1\right)+\frac{\So}{4}\left(\frac{u^2}{\tg^2}-1\right)^2\bigg]\notag\\
&+\frac{B vf_1(\tk u,-1)}{2\alpha r_0^3u}
+\frac{\tg^2(\Sp+\So)\Lambda_0}{\alpha r_0}\left(\frac{u^2}{\tg^2}-1\right)^2+\frac{4 B\Lambda_0}{\alpha^3r_0^3}.\label{eq:psitot}
\end{align}

We remind that $u$ and $v$ are not independent variables. Indeed, the volume of the current configuration is fixed by the growth, so that (see equation \eqref{eq:currvol})
\begin{equation}
u^2(v +\tfrac43 \Lambda_0u)=1+\tfrac43 \Lambda_0,
\end{equation}
where $\Lambda_0$ represents the reference aspect ratio.

Eq.~\eqref{eq:psitot} evidences that encasement, i.e.~the combined effect of bulk and coat growth, may not be simply renormalized as the effective growth of one of them (either bulk or coat) with the other considered as a simple elastic medium, so that encasement does not in general reduce to the constraint of confinement. For instance, in the present case, we describe the surface growth in terms of the renormalized parameters $\tb$ and $\tg$, but the bulk growth parameter $\alpha$ enters also the ratios between surface and bulk elastic constants. Therefore, when $\alpha$ (i.e. the bulk) grows, even if the relative growth parameters $\tb$ and $\tg$ remain constant, the surface effects become less important, affecting the stability thresholds and possibly their existence.

The straight shape is always a stationary configuration for the energy functional, because, setting $\Psi_0=\Psi\big|_{\tk=0}$:
\begin{align}\label{eq:psi2}
\Psi&=\Psi_0+\Bigg[\frac{\mu u^\frac83v^\frac13(7v^2-u^2)}{18}+\frac{\Sp u^2v^2(3v^2-\tb^2)\tg}{2\alpha r_0\tb^3}+\frac{B u v}{4\alpha r_0^3}\Bigg]\tk^2+o(\tk^2)
\end{align}
as $\tk\to0$, with
\begin{align}
\Psi_0&=\mu(u^\frac23v^\frac73+2u^\frac83v^\frac13-3u^2v) + \frac{\So\tb(u^2 - \tg^2)^2}{2\alpha r_0\tg^3}+\frac{\Sp\tg(v^2 - \tb^2)^2}{2\alpha r_0\tb^3}\notag\\
&+\frac{B v}{2\alpha r_0^3 u}+\frac{2r_0}{h_0}\left(\frac{(\So+\Sp)(u^2-\tg^2)^2}{2\alpha r_0\gamma^2}+\frac{2B}{\alpha^3  r_0^3}\right).\label{eq:psi0}
\end{align}

To check the stability of the straight shape we assume that the unknown curvature $\tk$ is small enough that $u$ may be determined by minimizing the zero-curvature strain energy $\Psi_0$, as defined in \eqref{eq:psi0}. We then replace the values obtained in the square-brackets term in \eqref{eq:psi2}, checking its sign to test stability. This leads to a number of possibilities for the bending instability of the growing encased cylinder, depending on the imposed material parameters.

\begin{figure}
\begin{center}
\includegraphics[width=8cm]{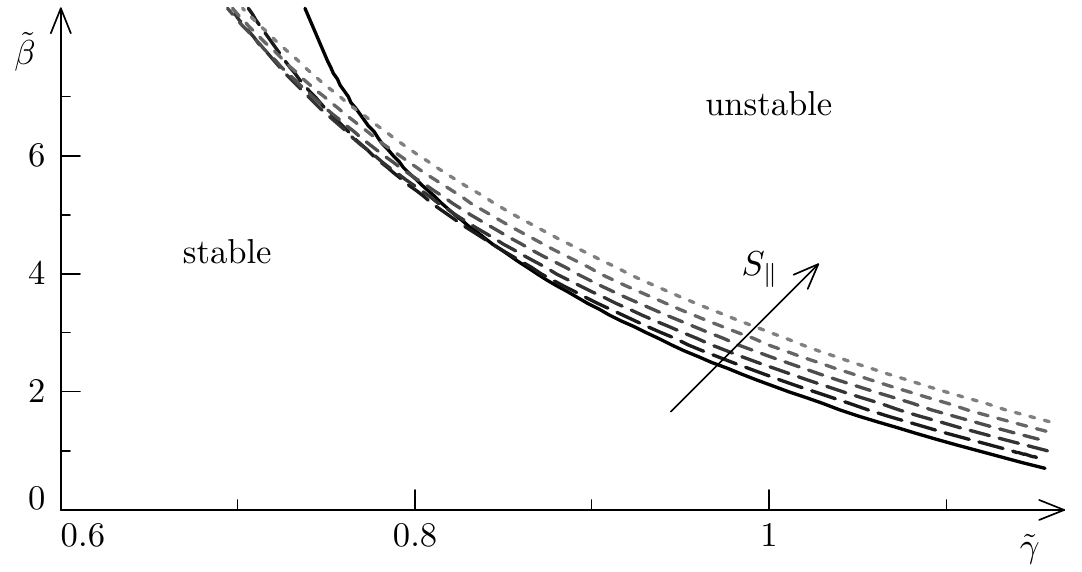}
\end{center}
\caption{Phase diagram evidencing the onset of an instability for the straight configuration when the longitudinal growth, represented by $\tb$, is large enough. Displayed values of the longitudinal-to-tangential stiffness ratio are, top to bottom, $S_\parallel/S_\perp=0.1,0.15,0.2,0.25,0.3,0.35$, with fixed $\So$.}
\label{fig:betacr}
\end{figure}

Among these possibilities, we only examine here a particular case. We look for the equilibrium configurations of a cylinder with initial aspect ratio $\Lambda_0=r_0/h_0=1$, with a soft (though incompressible) bulk, almost negligible bending stiffness and an anisotropic coat, reinforced on the tangential direction. Precisely we set $B / (2\alpha r_0^4\mu) = 10^{-3}$, $S_\perp / (2 \alpha r_0^2 \mu) = 10^2$, and vary $S_\parallel$, as indicated in Fig.~\ref{fig:betacr}. For each value of $S_\parallel$ and $\tg$ we determine numerically the optimal value of $u$ in \eqref{eq:psi0}, and check the straight stability sign in \eqref{eq:psi2}. Fig.~\ref{fig:betacr} evidences how, for each $\tg$, a critical value of $\tb$ exists, above which the straight configuration becomes unstable. Due to the meaning of the growth constants $\tg$ and $\tb$, this in turn indicates the critical value of the aspect ratio of the grown cylinder at which there is the onset of the bending instability. We notice in Fig.~\ref{fig:betacr} that such critical value increases as $\Sp$ approaches $\So$, indicating that surface elastic anisotropy helps triggering the instability of the straight configuration.

\begin{figure}
\begin{center}
\includegraphics[width=9cm]{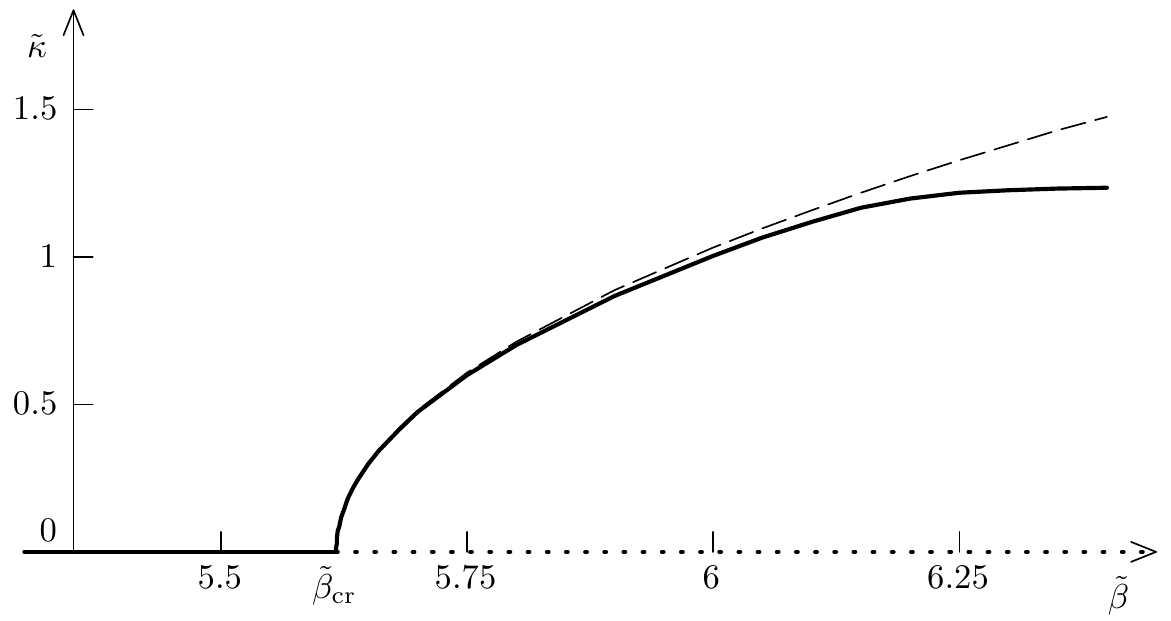}
\end{center}
\caption{Optimal value of the curvature $\tk$ as a function of $\tb$ for the material parameters provided in the text. The critical value of the longitudinal growth is $\tb\doteq 5.617$. The dashed line is a best fit for the critical behavior, showing that $\tk_\text{opt}\sim\sqrt{\tb-\tb_\text{cr}}$ as $\tb$ approaches the critical value leading to the bending bifurcation.  The same parameters as in Fig.~\ref{fig:betacr} are used.}
\label{fig:kappa}
\end{figure}

The features of the bifurcation connected with such instability may be analyzed by using the energy functional \eqref{eq:psitot}, obtaining for each value of $\tg$ and $\tb$ the optimal value of the axis curvature $\tk$.  We use the same material parameters as in Fig.~\ref{fig:betacr}, along with $S_\parallel/S_\perp=10^{-1}$, and $\tg=0.8$. We see from Fig.~\ref{fig:kappa} that the bending transition is of the second order, with the curvature continuously departing from zero (the underlying bifurcation for the growing encased cylinder corresponds to an axially-symmetric supercritical pitchfork).

\section*{Acknowledgements}

This research has been supported by the Italian Ministry of University and Research Grant
No.~200959L72B\_004, \emph{``Mathematics and Mechanics of Biological Assemblies and Soft Tissues''}. The Authors thank Donato Chiatante for many conversations and suggestions on morphogenesis in vegetables.

\appendix

\section{Geometry of bent cylindrical deformations}\label{sec:appbend}

\subsection{Bulk deformation}

Let us consider the reference and current illustrated in Fig.~\ref{fig:cylinders}:
\begin{align}\label{app:refconf}
\cB_0&=\big\{P\in\cE:P=O+\rho\cos\Theta\,\bex+\rho\sin\Theta\,\bey+z\,\bez,\notag\\
&z\in(-r_0,h_0+r_0),\,\rho\in\big[0,R_{r_0,h_0}(z)\big),\,\Theta\in[0,2\pi)\big\},\\
\cB_{\kappa}&=\big\{P\in\cE:P=Q_\kappa+\rho\cos\Theta\,\bex+\big(\rho\sin\Theta-\kappa^{-1}\big)(\cos\kappa z\,\bey-\sin\kappa z\,\bez),\!\!\notag\\
&\phantom{=\big\{}z\in[-r,h+r],\,\rho\in[0,R_{r,h}(z)],\,\Theta\in[0,2\pi)\big\},
\end{align}
where $r_0$, $h_0$, $r$, $h$, and $\kappa$ are positive parameters, $O$ and $Q_\kappa=O+\kappa^{-1}\,\bey$ are points in the three-dimensional Euclidean space $\cE$ with orthogonal basis $\{\bex,\,\bey,\,\bez\}$, and $R_{r,h}\colon (-r,h+r)\to \mathbb{R}^+$ is defined as
\begin{equation}\label{app:r0z}
R_{r,h}(z)=
\begin{cases}
\sqrt{r^2-z^2} & \text{if } z\in(-r,0]\\
r & \text{if } z\in(0,h)\\
\sqrt{r^2-(z-h)^2} & \text{if } z\in[h,h+r).
\end{cases}
\end{equation}

In all our calculations we consider the smooth deformation $\bchi:\cB_0\to\cB_{\kappa}$ such that, for every $P=O+X\,\bex+Y\,\bey+Z\,\bez\in\cB_0$, the coordinates of the transformed point $\bchi(P)=O+x\,\bex+y\,\bey+z\,\bez$ are given by
\begin{align}
x &= \frac{rX}{r_0}\notag\\
y &= \frac{rY}{r_0}+\Big(\kappa^{-1}-\frac{rY}{r_0}\Big)\Big(1-\cos\frac{\kappa hZ}{h_0}\Big) \label{app:defcoor}\\
z &= \frac{hZ}{h_0}+\Big(\kappa^{-1}-\frac{rY}{r_0}\Big)\sin\frac{\kappa hZ}{h_0}-\frac{hZ}{h_0}.\notag
\end{align}
We introduce the orthogonal unit vectors
\begin{equation}
\bffa(Z)=\cos\frac{\kappa h Z}{h_0}\,\bey-\sin\frac{\kappa h Z}{h_0}\,\bez\quad
\text{and}\quad\bfpa(Z)=\sin\frac{\kappa h Z}{h_0}\,\bey+\cos\frac{\kappa h Z}{h_0}\,\bez
\end{equation}
which, along with $\bex$, constitute an orthogonal basis in $\cE$. The deformation gradient associated with \eqref{app:defcoor} may thus be written as
\begin{equation}\label{app:bulkF}
\begin{aligned}
\bF&=\frac{r}{r_0}(\bex\otimes\bex+\bffa(Z)\otimes\bey)+\frac{h}{h_0}\left(1-\frac{\kappa r Y}{r_0}\right)\bfpa(Z)\otimes\bez.
\end{aligned}
\end{equation}
The isochoric deformation gradient is defined by
\begin{equation}
\mbF=J^{-\frac13}\bF, \quad \text{with}\quad J=\det\bF=\frac{hr^2(r_0-\kappa r Y)}{h_0r_0^3},
\end{equation}
and the related isochoric right Cauchy-Green strain $\mbC$ is
\begin{align}
\mbC&=\mbF^\top\mbF=\Big(\frac{h_0 r}{h(r_0-\kappa  rY)} \Big)^{2/3} \big(\bI-\bez\otimes\bez\big) + \Big(\frac{h(r_0-\kappa  r Y)}{h_0 r}\Big)^{4/3} \,\bez\otimes\bez.
\end{align}
In particular,
\begin{equation}\label{app:trmbC}
\tr\mbC=2\Big(\frac{h_0 r}{h(r_0-\kappa  rY)} \Big)^{2/3} + \Big(\frac{h(r_0-\kappa  r Y)}{h_0 r}\Big)^{4/3}.
\end{equation}

We end this computation by determining the condition on the parameters $h,r$ which ensures that the current placement $\cB_{\kappa}$ occupies a prescribed volume $V_\text{curr}=\alpha^3V_0$, where $V_0=r_0^2(h_0+\tfrac43 r_0)$ is the reference volume, and the positive scalar $\alpha$ provides the bulk volume increase. The current caps are two hemispheres of radius $r$, and so their volume is $\frac43\pi r^3$. The volume of the bent cylindrical part in $\cB_{\kappa}$ can be computed as
\begin{equation}
\begin{aligned}
\int_0^{h_0}dz\int_0^{r_0}\rho\,d\rho\int_0^{2\pi}d\Theta \;
\frac{hr^2(r_0-\kappa r \rho\sin\Theta)}{h_0r_0^3}=\pi hr^2
\end{aligned}
\end{equation}
and does therefore not depend on the curvature $\kappa$. As a consequence, the volume grows by a factor of $\alpha$ if and only if
\begin{equation}\label{eq:consvol}
r^2(h+\tfrac43 r) = \alpha^3 r_0^2(h_0+\tfrac43 r_0).
\end{equation}

\subsection{Surface deformation}\label{sec:appsurf}

We now compute the surface deformation gradient which maps tangent vectors on the reference coat $\partial\cB_0$ into tangent vectors on the boundary of the current deformed body $\partial\cB_{\kappa}$. We refer to \cite{99stog} for more details. We first consider the deformation which maps the cylindrical part of $\partial\cB_0$ into the corresponding portion of $\partial\cB_{\kappa}$. We introduce the notations:
\begin{equation}
\begin{aligned}
\ber(\Theta)&=\phantom{-}\cos\Theta\,\bex+\sin\Theta\,\bey\\
\bet(\Theta)&=-\sin\Theta\,\bex+\cos\Theta\,\bey\\
\bffb(Z)&=\cos\big(\kappa h Z/h_0\big)\bey-\sin\big(\kappa h Z/h_0\big)\bez\\
\bfpb(Z)&=\sin\big(\kappa h Z/h_0\big)\bey+\cos\big(\kappa h Z/h_0\big)\bez,
\end{aligned}
\end{equation}
and identify points on $\partial\cB_0$ by the coordinates $(\Theta,Z)\in\{[0,2\pi)\times[0,h_0]\}$, with $\bY(\Theta,Z)=O+r_0\,\ber(\Theta)+Z\,\bez$. Then, $\by(\Theta,Z)=\bchi(\bY(\Theta,Z))$ is given by
\begin{equation}
\by(\Theta,Z)=Q_\kappa+r\cos\Theta\,\bex+\big(r\sin\Theta-\kappa^{-1}\big)\bffb(Z),
\end{equation}
where again $Q_\kappa=O+\kappa^{-1}\,\bey$. The tangent unit vectors in $\partial\cB_0$ and $\partial\cB_{\kappa}$, which in this simple example may be identified with their duals, are respectively given by
\begin{equation}
\begin{aligned}
\bT_\Theta&=\bT^\Theta=\bet, &&\bt_\Theta=\bt^\Theta=-\sin\Theta\,\bex+\cos\Theta\,\bffb(Z)\\
\bT_Z&=\bT^Z=\bez, &&\bt_Z=\bt^Z=\bfpb(Z),
\end{aligned}
\end{equation}
thus leading to the following grown and current normal unit vectors
\begin{align}
\bN&=\be_r\qquad\qquad
\bn=\cos\Theta\,\bex+\sin\Theta\,\bffb(Z).
\end{align}
The metrics associated with the parameterizations provide the local area dilation factor, which connects the current and the reference area element: $da=\fJ(\Theta)\,dA$, with
\begin{equation}\label{app:jacarea}
\fJ(\Theta)=\frac{hr(1-\kappa r \sin\Theta)}{h_0 r_0}.
\end{equation}
The surface deformation gradient
\begin{align}
\bA&=\frac{r}{r_0}\,\bt_\Theta\otimes\bT^\Theta+
\frac{h(1-\kappa r\sin\Theta)}{h_0}\,\bt_z\otimes\bT^z,
\end{align}
may be decomposed into an elastic and a growth component: $\bA=\bAe\bAg$. This latter represents an anisotropic elongation of factors $\beta,\gamma$, so that
\begin{align}
\bAg&=\gamma\,\bT^\Theta\otimes\bT^\Theta+
\beta\,\bT^z\otimes\bT^z,\qquad \text{and}\\
\bAe&=\bA\bA_\text{g}^{-1}=\frac{r}{\gamma r_0}\,\bt_\Theta\otimes\bT^\Theta+
\frac{h(1-\kappa r\sin\Theta)}{\beta h_0}\,\bt_z\otimes\bT^z.
\end{align}
The elastic strain tensor is then
\begin{align}\label{app:isostrain}
\bEe&=\tfrac12\big(\bAe^\top\bAe-\bIs)\notag\\
&=\frac12\left[\left(\frac{r^2}{\gamma^2 r_0^2}-1\right)\,\bT^\Theta\otimes\bT_\Theta+
\left(\frac{h^2(1-\kappa r\sin\Theta)^2}{\beta^2 h_0^2}-1\right)\,\bT^z\otimes\bT_z\right],
\end{align}
where $\bIs=\bT^\Theta\otimes\bT^\Theta+\bT^z\otimes\bT^z$ is the identity tensor in the reference tangent plane.

We obtain the mean curvature $H$ from the current tensor curvature
\begin{align}\label{app:H}
H=\tr\left(-\frac{1}{r}\,\bt^\Theta\otimes\bt^\Theta
-\frac{\kappa \sin\Theta}{1-\kappa r\sin\Theta}\,\bt^z\otimes\bt^z\right)=-\frac{1}{2r(1-\kappa r\sin\Theta)}.
\end{align}

The deformation in the caps is simpler, as it transforms a reference half-sphere of radius $r_0$ into the current half-sphere of radius $r$. Since $\bAg|_\text{caps}=\gamma\bIs$, we obtain
$da=\fJ\,dA$, with
\begin{equation}
\fJ\big|_\text{caps}=\frac{r^2}{r_0^2},\quad
\bAe\big|_\text{caps}=\frac{r}{\gamma r_0}\,\bIs,\quad \text{and}\quad \bEe\big|_\text{caps}=\frac12\left(\frac{r^2}{\gamma^2 r_0^2}-1\right)\,\bIs.
\end{equation}
The curvature tensor is isotropic, with mean curvature
\begin{equation}\label{app:Hcaps}
H\big|_\text{caps}=-r^{-1}.
\end{equation}

\end{document}